\documentclass[12pt]{revtex4}
\usepackage{graphicx}

\begin{document}

\title{Local Quantum Reality}
\author{Vlatko Vedral}
\affiliation{Clarendon Laboratory, University of Oxford, Parks Road, Oxford OX1 3PU, United Kingdom
\\Centre for Quantum Technologies, National University of Singapore, 3 Science Drive 2, Singapore 117543\\
Department of Physics, National University of Singapore, 2 Science Drive 3, Singapore 117542}

\begin{abstract}
Unperformed measurements have no results. Unobserved results can affect future measurements.
\end{abstract}

\maketitle

Quantum physics and special relativity are unified into quantum field theory. The need for fields is there because relativity rests on the principle of locality, namely the fact that there is no action-at-a-distance in nature. In other words, actions in a localised spatial region cannot have an instantaneous effect in another region of space. Of course, special relativity says more than that, namely that no physical influence can propagate faster than light, but the exact maximum speed of propagation is not relevant for locality as long as this speed is finite. The need for quantum features (in quantum field theory) is there because nature does not allow us to specify all the properties of these fields simultaneously in the same way as in the classical case (as numbers). For instance, classically, the electromagnetic field requires six numbers (three components of the electric and three components of the magnetic field) to be specified at each point in space and at every instance of time. Quantumly, the electric and the magnetic fields can also be specified at each point and at every time, however they are no longer allowed to be numbers! Instead, they are operators that generally do not commute. This of course is due to the fact that the Uncertainty Principle in quantum physics prohibits us from assigning numbers to all the components of the field at the same time. Locality is still satisfied because no matter how you change these operators in one region, they will not change instantaneously anywhere else. So, fields are there because of locality (no action at a distance, relativity) but they are fields of operators (because of their quantum nature) and not numbers as they would be in classical physics.  

Merging quantum physics and special relativity into quantum field theory has given us a number of non-trivial and surprising consequences \cite{Wilczek}: the need for a creation and annihilation of particles, the existence of anti-particles, the existence of the quantum vacuum and its effects, an explanation for the existence of fermionic and bosonic particles and the indistinguishability of identical particles to name the main ones. Quantum field theory is also our most accurate description of nature to date. 

Now, however, I would like to elaborate on the need for operators as elements of reality and this brings me to the abstract of my article. I have borrowed it from two of my colleagues. The first sentence is from a paper of Asher Peres, published in the American Journal of Physics a while ago \cite{Peres}, explaining the notion of quantum entanglement and Bell’s inequalities. The meaning of the phrase “unperformed measurements have no results” reflects exactly the fact that the elements of reality in quantum physics are operators and not numbers. So, since we think of measurement outcomes as being numbers which give us the magnitude of the measured quantity, the fact that there are no local hidden variables (no locally assigned numbers) that can account for the correlations exhibited by entangled states, tells us that these numbers do not exist independently of us making measurements. One sometimes hears it said that therefore reality does not exist according to quantum mechanics. I believe that this is a semantic issue. It is more precise to say that reality does exist, but that it is a quantum reality of the underlying operators, and not the classical reality of numbers. In that sense quantum field theory is a local (quantum-)realistic theory. 

So far so good. But things would not be that interesting if we stopped here. 

The second sentence of my abstract is a (slightly rephrased) sentence from a recent paper of Anthony Sudbery  \cite{Sudbery}, who has, along with many others, extensively discussed the implications of quantum entanglement for our understanding of the measurement process.  The phrase “unobserved results can affect future measurements” sounds like a Zen Koan \cite{Koan}. At first sight it appears contradictory. If a result is not observed, but a measurement has been made, that ought to mean that a different result has been observed instead. So how can something that is not observed still be relevant in the future? What is the sound of one hand clapping? 

Quantum physics has since its inception been frequently compared with the Eastern mystical traditions like Zen. However, no matter what one thinks of analogies like that, there is one aspect I believe Zen and quantum physics certainly do have in common: they both emphasise the basic non-duality of nature. And here I do not mean this in relation to the so-called wave-particle duality (which is of historical importance, but no more than that, really; quantum fields are the fundamental objects and particles are just excitations of the modes of these fields). Instead, what I have in mind is that, if quantum physics is to describe everything in the universe, then the distinction between the observers and the observed must vanish. If, by assuming the universality of the quantum, observers are also fully subject to the quantum laws (and there is currently no reason to believe otherwise), then a measurement process simply creates entanglement between the observers and the observed. And there is nothing more to it. Full stop.

Entanglement is a symmetric property of two entangled systems. The observer is as entangled to the observed as the observed is to the observer. The two roles are completely interchangeable as far as entanglement (and therefore measurements) are concerned. Expressed in terms of the quantum elements of reality we talked about earlier, there are operators pertaining to the observer that become correlated with (entangled to) the operators pertaining to the observed system when a measurement is performed. But these operators are no mere numbers as we said. They contain much more information. They contain information about all possible measurements of all possible observables. It is for this reason that the unobserved outcomes can have future effects. 

There is a host of related experiments that would perfectly demonstrate this point (otherwise, of course, we would not be talking about it). I described one of them in my two recent articles “Observing the observer: I and II” \cite{Vedral}. Variants have also been outlined by many other physicists, such as Sudbery himself, then David Deutsch \cite{Deutsch} and David Albert \cite{Albert} to name a few.  I will briefly sketch the key thought experiment here just for completeness. 

Imagine that a photon (which is an excitation of a mode of the quantum electromagnetic field) impinges on Bob’s sunglasses (which could also be described in quantum field theory, but let’s not complicate things too much without necessity). After the interaction with the surface of sunglasses, the photon is in a superposition of being transmitted and being reflected. When transmitted, the photon enters Bob’s eye, thereby generating a nerve impulse that ultimately results in Bob seeing it (it’s an intricate process to be sure, but all I need to assume is that it is quantum). However, when reflected the photon just bounces back and does not generate any neurological response in Bob’s brain (this will be the unobserved result, although, strictly speaking, it is still an observation, but an observation of no photons). So, Bob and the photon are entangled: in one branch, the photon has been transmitted through the glasses and has stimulated something in Bob’s brain, while, in the other branch, the photon is reflected and Bob is left unperturbed. Now, in the first branch Bob “knows” he has seen the photon, while in the second one he “knows” he has not. This of course is the good old Schrödinger’s cat state (between Bob and the photon) and, without too much exaggeration, it is fair to say that all quantum experiments are really just more or less complicated versions of Schrödinger’s (less complicated because we are still not able to interfere cats or humans, however, more complicated in terms of the degree of entanglement between different constituents). 

I will now explain how the Bob that has seen the photon will be affected by not seeing the photon (i.e. by the Bob in the other branch of the state) and vice versa. The idea is, as always, to interfere the two possibilities. This is in practice difficult beyond belief, however, if quantum physics is universal, it ought to be possible with sufficient resources. For this purpose, we imagine another physicist, Alice, who controls the experiment where Bob and the photon become entangled. Alice now needs to reverse the entangling process that has resulted in Bob’s observation (and non-observation in the other branch). If the process is perfectly reversed, then Bob and the photon are in their initial states, before the photon has interacted with the sunglasses and before Bob has made any observation. This means that both alternatives must have existed and were perfectly superposed in an entangled state, for otherwise the final state would not be the same as the initial one (this is what we mean by interference: a superposition is first created and is then subsequently reversed). This also means that both outcomes in both branches must have existed within the entangled state of Bob and the photon. The elements of reality encoded into quantum operators are there all along, independently of what has been observed and what has not been observed.
And what I have just described here is simply a version of the Mach-Zehnder interferometer where Bob undergoes interference along with the photon. 

Of course, Alice and Bob need not be sentient beings. They could be automata (humans probably are, as Thomas Henry Huxley famously maintained, just sophisticated automata \cite{Huxley}, but this is irrelevant for the arguments here), preprogramed to undergo certain interactions. Consciousness plays no role. Alice is as quantum as Bob, and both as quantum as the photon (albeit much more complex in the sense of requiring more operators to describe them than a single photon). We could have more observers/systems in the experiment (as Eugene Wigner imagined \cite{Wigner}), and the conclusions would still be the same.  

The philosophy based on quantum physics is frequently presented as idealistic, implying that there is no reality out there beyond the one that we create through measurements (which is when we obtain real numbers as experimental results). Or, as some physicists would say, if we want reality (again in the sense of real numbers pre-existing as a local underlying description) then we have to give up the notion of locality. This is why quantum entanglement is sometimes referred to as quantum non-locality. But this ought to, strictly speaking, be called the Bell-type non-locality (i.e. measurements at one place must affect instantaneously another place if the underlying reality is based purely on numbers) in order to discriminate it from the non-locality we discussed and which is successfully avoided in quantum field theory. 
 
One final remark. There is another effect, called the Aharonov-Bohm effect \cite{AB}, which is also frequently claimed to be non-local. An electron, undergoing interference in a magnetic field that is localised in a small region far away from where the electron is (and is otherwise zero), does actually get affected by this field. If we think in terms of local quantum operators that pertain both to the electron as well as to the electromagnetic field, then this effect is also completely local (like all other phenomena captured by quantum field theory). Namely, the electromagnetic field always couples to the electron locally (through the quantum operators) and then propagates at the speed of light to and from the localised magnetic field. This is something that Chiara Marletto and I have recently described in a publication in Physical Review Letters \cite{MV}.  

I hope to have convinced you that we should think about quantum reality as follows. The reality is out there, it is local and independent of us, but it is not underpinned by numbers (that’s the price we need to pay, if you like that kind of a language, in order to make sense of all our observations). This corresponds exactly to how we think of quantum fields and, if you think of everything in the universe as made up of fields, it yields perfectly consistent, although frequently counter-intuitive results (this is true no matter what interpretation of quantum physics one happens to support). It offers us a mind-blowingly accurate picture of the universe that is at present only obscured by one cloud. Gravity. But methinks this is only a matter of time…and only God knows what the resulting elements of reality will be after gravity has finally been quantized (on second thought, maybe not even God, given that reality is, judging by what we currently know, likely to remain quantum).

\textit{Acknowledgments}: The author is grateful to Chiara Marletto for comments on this work. He acknowledges funding from the National Research Foundation
(Singapore), the Ministry of Education (Singapore), the Engineering and Physical Sciences Research Council (UK), and Wolfson College, University of Oxford.

\end{document}